\documentclass[journal]{IEEEtran}
\ifCLASSINFOpdf
\else
\fi
\usepackage{lineno,hyperref}
\usepackage{subcaption}
\usepackage{multicol}
\usepackage{amssymb}
\usepackage{amsmath}
\usepackage{txfonts}
\usepackage{mathdots}
\usepackage{graphics} 
\usepackage{graphicx}
\usepackage{pgfplots}
\usepackage{makecell}
\usepackage{multirow}
\usepackage{tabularx}
\usepackage{algorithm}
\usepackage[noend]{algpseudocode}

\begin{document}
%
\title{An Ensemble Learning Based Classification of Individual Finger Movement from EEG}
%
%
%

\author{Sutanu~Bera,
        Rinku~Roy, Debdeep~Sikdar, and Manjunatha~Mahadevappa
\thanks{Sutanu Bera, Debdeep Sikdar, Manjunatha Mahadevappa are with the School of Medical Science and Technology, Indian Institute of Technology Kharagpur, India. \scriptsize email: sutanu.bera@iitkgp.ac.in, deep@iitkgp.ac.in, mmaha2@smst.iitkgp.ac.in}
\thanks{Rinku Roy is with Advanced Technology Development Centre, Indian Institute of Technology Kharagpur, India. \scriptsize email: rinku.roy@iitkgp.ac.in}
}

\maketitle

\begin{abstract}
Brain computer interface based assistive technology are currently promoted for motor rehabilitation of the neuromuscular ailed individuals. Recent studies indicate high potential of utilising electroencephalography (EEG) to extract motor related intentions. 
Limbic movement intentions are already exhaustively studied by the researchers with high accuracy rate. But, capturing movement of fingers from EEG is still in nascent stage. 
In this study, we have proposed an ensemble learning based approach for EEG in distinguishing between movements of different fingers, namely, thumb, index, and middle. Six healthy subjects participated in this study. Common spatial patterns (CSP) were extracted as features to classify with the extra tree or extremely randomized tree binary classifier. The average classification accuracy of decoding a finger from rest condition was found to be $74\%$, wheres in discriminating of movement of pair of fingers average accuracy was $60\%$. Furthermore, error correcting output coding (ECOC) was added to the binary classifier to use it in multiclass classification. The proposed algorithm achieved a maximum kappa value of 0.36 among the subjects.

\end{abstract}

%
\IEEEpeerreviewmaketitle

\section{Introduction}
Holding small and lightweight objects by two opposing fingers is called \textit{Pinch Grasp} \cite{feix2016grasp}. Such objects encompass from pens to spoons, from keys to coins, even from salt to paper etc. In most of the cases, thumb remains one of the fingers in opposition, whereas other finger is appropriated by our brain based on the demand of the situation. However, neuromuscular disorders hinder the individuals to control the movements of limbs and fingers as well. Although mechanically controlled prosthetic hand help them to open/close their hand, but finger manipulation for a stable grip is still missing. Brain controlled prosthesis is a solution to bypass the damaged neuromuscular pathway altogether. 


There have been exhaustive studies executed to establish relationship between EEG and movements in voluntary body parts like wrists \cite{gu2009single}, upper limbs \cite{doud2011continuous}, elbows and shoulders \cite{zhou2009eeg}, legs \cite{pfurtscheller2006mu}, and tongue \cite{morash2008classifying}. However, finer manipulation of the fingers is still at nascent stage. Although fingers being the most dexterous part of our body to play an irreplaceable role in object grasping, the extent of non-invasive recording techniques to provide information for movement decoding is still unexplored. ECoG recordings in humans \cite{acharya2010electrocorticographic, kubanek2009decoding, miller2009decoupling} and single unit recordings in monkeys \cite{aggarwal2009cortical, hamed2007decoding} have shown the potential to detect movements of individual fingers. 
The muscle mass involved in finger movements is smaller than in limb or hand movements and neuronal discharges of motor cortex neurons are correspondingly smaller in finger movements than in arm or wrist movements \cite{pfurtscheller2003spatiotemporal}, which makes them difficult to detect. Another potential challenge is the spatially overlapping finger representation in somatosensory cortex. Penfield and Boldrey found that the distance along the Rolandic fissure that evoked finger movements after cortical stimulation was 55 mm \cite{penfield1937somatic}.

Spatio-spectral features from ECoG-based BCI studies in individual finger movements have shown remarkable results \cite{zanos2008electrocorticographic, flamary2012decoding, liang2012decoding, kubanek2009decoding, shenoy2007finger, onaran2011classification, samiee2010five, wang2009human, miller2009decoupling, shenoy2007finger, onaran2011classification, samiee2010five} and micro-ECoG grid recordings \cite{wang2009human}. In one study \cite{miller2009decoupling}, increase in broadband (up to 200 Hz) spectral power and decrease in
characteristic spectral power were reported in both $\alpha$ (8-12 Hz) and $\beta$ (13-30 Hz) bands during individual finger movements from one hand, in which the broadband phenomenon has been demonstrated sensitive to movements performed by different fingers. These ongoing ECoG studies have demonstrated the feasibility of decoding individual finger movements using electrical potentials generated by the human brain, inspiring research in such decoding tasks using noninvasive EEG. EEG has been used by the researchers to decode different grasping actions \cite{agashe2015global, agashe2013decoding, erfanian1999prediction}. Strong correlation between actual grasping and corresponding motor imagery has been already established \cite{8008384}. Various linear and non-linear features were utilised to decode not only the grasp aperture selection but also discrimination among various power grasps and precision grasps imagination \cite{Roy2018}. Results suggested changes in event-related desynchonization (ERD) among $\alpha$ and $\beta$ band of EEG during thinking of holding an object. Although classification of \textit{Pinch Grasp}, \textit{lateral Grasp}, and \textit{Tripod Grasp} were reported in the study, the finger selections for a particular precision grasp is still unexplored. Few studies also suggested the possibilities of decoding individual finger movements from EEG also \cite{quandt2012single, blankertz2006berlin}. One of these studies discussed the decoding of four fingers movements (no ring finger) using EEG and suggested the EEG recording was not very robust in part due to its low spatial resolution \cite{quandt2012single}. 
In another study, all the individual finger movements ( thumb, index, middle, ring, little) were decoded from one hand \cite{liao2014decoding}. In all these studies, individual finger movement decoding problem was solved by two-class classification problem i.e., EEG data from each finger was compared with another finger. Selection of individual finger movement from multi-class problem is yet to be explored for grasping task.

The goal of this study is to develop a multi-class classifier for decoding individual finger movement. The EEG data was captured from six healthy subjects with a finger movement protocol. In this study, we focused only in thumb, index and middle as they were mostly used for pinch grasp. The raw data were first chopped into several frequency band with each of 2 Hz limit. The results depicted that in most of the times, the ERD occurred in $\alpha$ band frequency range. Due to poor SNR of EEG generated for finger movement, Common spatial Pattern (CSP) used here to achieve better SNR by maximizing the difference of two data populations. To solve the multi-class problem, we started with the binary classification problem. An ensemble learning technique, Extra-tree classifier used here to classify between two sets of finger data. The extra tree classifier was extended to solve the multi-class problem. \textit{Error correcting Output coding (ECOC)} method is used here with the combination of Extra-tree classifier to develop an advanced multi-class classifier. The result suggests that the combination of ECOC with extra tree classifier will provide a new dimension towards BCI experiment.  

\section{MATERIALS AND METHODS}
\subsection{Subjects}
In this study, six healthy subjects volunteered aged between 23-28 years. They were right handed and having not neuromuscular disability. The participants were asked to perform finger movement as per the visual cue with their dominating arm. To avoid any biasness toward the expected stimulus beforehand, the visual sequences in the cue were randomised. This study protocol was approved by the Institution Ethical Committee (IEC) at the Indian Institute of Technology Kharagpur, Kharagpur. This research adhered to the tenets of the Declaration of Helsinki. Written consents were taken from the subjects before participating in the experiment. 

\subsection{Experimental Setup}
\begin{figure}[hbtp]
    \centering
    \includegraphics[width=.4\textwidth]{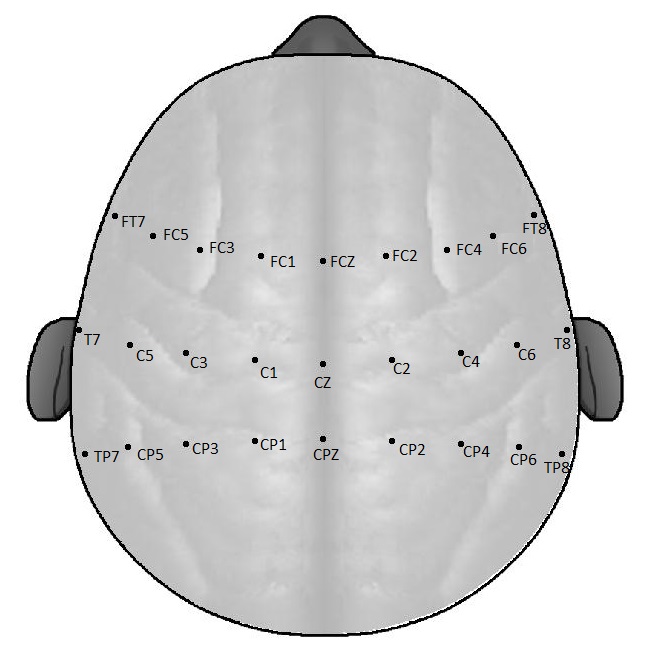}
    \caption{Electrode Positions used in this study}
    \label{montageEEG}
\end{figure}

Subjects were seated on a chair in front of a table where a computer screen was placed to show the visual stimulus to the subjects. Subjects' palm of the dominant hand was kept flat on the table. They were asked to restrict their eye movements and other muscle movements less as much as possible. EEG signals were recorded at a sampling rate of 512 Hz from g\textsuperscript{\textregistered}.Hiamp (g.tec, Graz, Austria) hardware with 27 monopolar channels viz. FT7, FC5, FC3, FC1, FCZ, FC2, FC4, FC6, FT8, T7, C5, C3, C1, CZ, C2, C4, C6, T8, TP7, CP5, CP3, CP1, CPZ, CP2, CP4, CP6, TP8 by scalp electrodes placed according to the International 10-20 system as shown in Figure \ref{montageEEG}. The reference electrode was placed on the right earlobe and the ground electrode (GND) was placed on the forehead (AFz). These electrode positions were chosen to cover the motor cortex area. 8th Order butterworth bandpass filter with 5- 100 Hz range was used to remove the low frequency noise. A 48-52 Hz filter were also configured in the acquisition software to remove linear trends and electrical noise.

\subsection{Task}

\begin{figure*}[t]
    \centering
    \includegraphics[width=\textwidth]{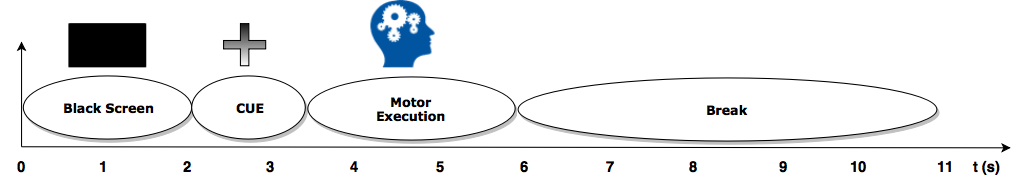}
    \caption{Experimental Protocol for EEG data collection}
    \label{StimulusTiming}
\end{figure*}

The experiment was performed in a sound proof room. 
The subjects were asked to keep their palm on the table at resting position throughout the experiment. They were only allowed to move their finger as per visual cue. Recording sessions were conducted in three consecutive days. For the first two days, the sessions were focused in training the subject to improvise the overall concentration of the subjects. Third day data were used for further processing. 
The experimental protocol started with a blank screen of 2 sec. The subjects were instructed to relax and concentrate on the task. A fixation cross was then shown to the subject for 1.5 sec followed by the motor execution task stimuli. One among three fingers (thumb, index, middle) was instructed to lift and hold above a certain level for 2.5 sec at a time and then lower it to resting position. The finger orders were shown to the subject in random sequence. Between consecutive visual cues, a blank screen of 5 sec was given to subject for \textit{rest} condition. The overall timeline of the visual stimulus is shown in Figure \ref{StimulusTiming}. 



\subsection{EEG Analysis}

\begin{figure*}[h]
    \centering
    \includegraphics[width=\textwidth]{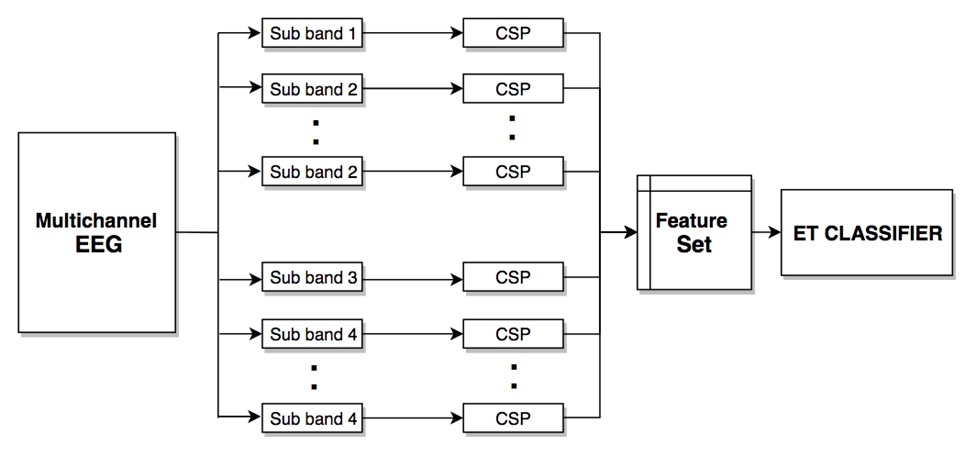}
    \caption{Proposed Algorithm for Binary Classification}
    \label{binaryalgo}
\end{figure*}

All the analysis were executed in open source python library SciPy, and Scikit-learn, under Windows 7 64bit environment in an Intel Core i5 Processor based system. Fig.~\ref{binaryalgo} shows process flow of the overall EEG analysis. 
Following re-referencing of the EEG data, it was segmented according to the cue timing. Ocular, muscular and cardiac artifacts were removed along with other noise followed by decomposition into several frequency subband. The algorithm design was initiated with binary classifier and ultimately it was extended for a multi-class scenario. In this study, four classes were considered as \textit{thumb}, \textit{index}, \textit{middle} and \textit{rest}. For each of the subbands, Common Spatial Pattern (CSP) were extracted. Linear Discriminant Analysis (LDA) was then employed to isolate subject specific optimal subbands affected by the assigned task to reduce the overall computational complexity. At testing phase, we have utilised only the selected frequency subbands by LDA to classify. Ensemble Extra-tree classifier was implemented on the selected data-set for binary classification. The classification was done in two stages: in fist stage, all individual finger movements were separated from the \textit{rest} condition. In the next stage, three pairwise combinations were chosen: Thumb vs. Index, Thumb vs. Middle, Index vs. Middle. The binary Extra-tree classifier was then summed up with Error correcting output coding (ECOC) as an extension to multi-class classification. 

\subsubsection{Frequency band decomposition}

The acquired EEG signals were decomposed into several frequency subbands each with 2Hz bandwidth starting from 5Hz. We have limited our study upto 40Hz resulting into 17 segments. Finite impulse response filter with Hamming window was employed to construct the filter.

\subsubsection{CSP method}

CSP \cite{ramoser2000optimal} technique was employed to improve the SNR of the filtered signal. The methodology was described in detail in our previous study \cite{8513421}.

\subsubsection{Appropriate frequency band selection using LDA} 

\begin{figure*}[h]
    \centering
    \includegraphics[width=\textwidth]{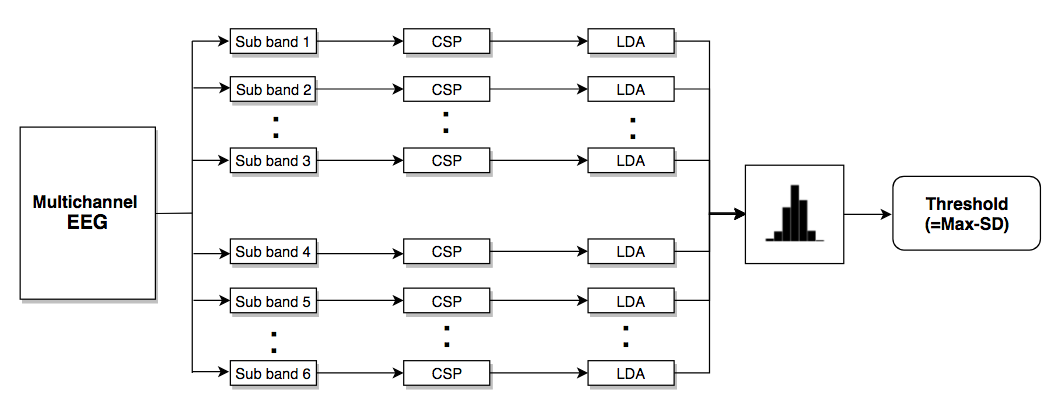}
    \caption{Frequency Band Selection Process}
    \label{frequencyb}
\end{figure*}

Among all the seventeen computed frequency bands, the appropriate frequency band should be selected for further processing. The choice of the frequency band mostly depends on the task i.e, for execution of different body parts, this frequency range should be different. Also this frequency range is little bit subject specific i.e., for different subject, frequency range can be different. For this reason, first and last two component of spatially filtered signal of every sub band is classified by LDA separately as shown in Figure \ref{frequencyb}. 5 fold cross-validation was used for the process. The classification accuracy of every band were assigned as the \textit{frequency score ($f$)} of that band. A Threshold value ($T_H$) was decided from that \textit{frequency score} using the following formula,

\begin{equation}
T_H = max(f)- Standard Deviation(f)\\
\label{eqn8}
\end{equation}

The frequency bands having \textit{frequency score} greater than the calculated threshold value ($T_H$)  were selected for the classification. The overall frequency selection method is described in Figure \ref{frequencyb}. This frequency selection procedure was appointed only during the training phase of the ensemble network. Only those selected frequency bands were used in the evaluation phases of the classification for increasing the computation time. 

\subsubsection{Binary Classification}

In the first stage, all the individual finger movement were distinguished from the \textit{resting} stage when fingers were laid down on the table top. After that pairwise classification of \textit{Thumb vs. Index, Thumb vs Middle and Middle vs Index} were done for checking whether they distinguished from each other or not.\\

\paragraph*{Extra Trees Classifier (ET)}

ET classifier is used here to predict an output from two binary classes. It operates on the idea of ensemble learning where trees are constructed for deciding the output class of given set of data \cite{geurts2006extremely, wehenkel2006ensembles}. In this technique, multiple unpruned decision trees were developed during learning and output class were decided by the 'mode' of all classes generated by the individual trees. Other ensemble learning techniques like Random forest classifier, ET avoids bootstraping concept. The most important point for concerning regarding ET is that nodes are here split by the optimal cut point which are generated randomly from the whole data set. \\

ET structure requires three input parameters for the classification- $K$, $n_min$, $M$. The optimal value of $K$, $n_min$, and $M$ were selected by cross validation method. ET first randomly selects $K$ attributes, $\{a_1,.....,a_K\}$ from a set of large data set. Using this attributes $K$ attributes, it randomly generates $K$ number of splits $\{s_1,....,s_K\}$ for the given data set. For better randomization procedure, a optimum value for the $K$ is required. Large value of the $K$, slows down the randomization process and also degrades the ability for correct classification. All the $K$ number of splits having a score value. The splits having highest score value were selected for the next stage. The minimum number of samples required for splitting a node, is also an important factor for the ET structure. The value of $n_min$, should be optimized for smaller variance and higher bias due to which classification accuracy is increased. Hence, its optimal value depends on the level of output noise on the data set. The last parameter, $M$, denotes the number of trees considered in the ensemble structure. As prediction error monotonically decreasing in each tree, a large no. of $M$ is better for the higher classification accuracy. Depending on the sample size and other computational requirements, $M$ is decided for a specific problem.

\subsubsection{Multi-class Extension of binary classifier}

Pair-wise classification has already been done in the previous stage. But finger selection during pinch grasp requires multi-class classification i.e., selecting a single finger at a time among multiple. Here, multi-class classification was done with the Extra Tree classifier in addition with Error correcting output coding (ECOC) approach. Other approaches like Pair Wise \cite{muller1999designing, duda2001pattern}, One-versus-Rest \cite{dornhege2004increase, duda2001pattern}, Divide-and-Conquer \cite{zhang2007algorithm, chin2009multi} has already been used for classification of various motor activities. In this study, ECOC \cite{dietterich1994solving} is used here with the combination of ET classifier.We named our technique as Randomised Ensemble leraning(REL) technique. In BCI domain, ECOC is still not used for classification purposes. Here, it was explored to achieve better classification accuracy. Thumb, Index, Middle finger and rest were the four classes considered here. 


\paragraph*{Error Correcting Output Coding (ECOC)}

ECOC used here to solve four class problem using binary classifiers. The output of each binary classifier was gone through several stages of ECOC procedure. In very first stage, output of all the binary classifiers was added up to generate a code word. Here, a 7 bit long \textit{codeword} was generated using \textit{Exhaustive coding Scheme}. According to that scheme, for solving a $k$ class problem, each $codeword$ consists $2^{k-1}-1$ bits. The \textit{Exhaustive code} representation for four class problem is shown in Table \ref{ecoc}. For class1, all the seven bits are ones. Class2 have $2^{k-2}$ zeros, followed by $2^{k-2}-1$ ones. Class3 consists of $2^{k-3}$ zeros, followed by $2^{k-3}$ ones, followed by $2^{k-3}$ zeros, followed by $2^{k-3}-1$ ones. In row $i$, there are alternating runs of $2^{k-i}$ zeros and ones. For each column of the code matrix, separate binary functions were learned. In the next stage of ECOC, generated $codeword$ for an unknown data set is compared with $codewords$ of all the classes. Distance between the obtained string from all the other classes were computed. $Hamming Distance$ was used here for measuring the distance between $codewords$. The unknown data is predicted to the class which have minimum $Hamming Distance$ from the generated $code word$.


\begin{table}[h]
\centering
\caption{Code Matrix for Four Class using Exhaustive code }
\label{ecoc}
\begin{tabular}{|c|ccccccc|}
\hline
class1 & 1 & 1 & 1 & 1 & 1 & 1 & 1 \\ \hline
class2 & 0 & 0 & 0 & 1 & 1 & 1 & 1 \\ \hline
class3 & 0 & 0 & 1 & 1 & 0 & 0 & 1 \\ \hline
class4 & 0 & 1 & 0 & 1 & 0 & 1 & 0 \\ \hline
\end{tabular}
\end{table}


\subsubsection{Proposed Algorithm}

\begin{algorithm*}
\caption{Proposed Multi-class Algorithm}\label{alecoc}
\begin{algorithmic}[1]
\Procedure{Training}{Multi-trial EEG of dimension ($n\times N\times T$), Code matrix $C$ of dimension ($p\times q$)}\Comment{$n$: number of trail, $N$: number of channel, $T$: Duration, $p$: number of class, $q$: code word length}
\For{j:=1 \textbf{to} $p$}
\begin{itemize}
    \item Generate two subsets  $S^{+}$ and $S^{-}$ from all training samples where 
    \begin{itemize}
        \item $S^{+}$ for $C_{ij}$ =1 
        \item $S^{-}$ for $C_{ij}$ =0
    \end{itemize}
    
    \item Select the optimal frequency band for this particular problem.
    \item Train the binary classifier $\lambda_{j}$ to discriminate $S^{+}$ from $S^{-}$ 
\end{itemize}
\EndFor
\State \textbf{return} $<\lambda_{1}, \lambda_{2}, ... \lambda_{p}>$
\EndProcedure
\Procedure{Evaluation}{EEG($N\times T$)}
\Repeat
\State  Predict output of unknown EEG trial using $\lambda_j$
\Until $j:=p$
\State Create the code word by joining the predicted class label from each $\lambda_{j}$.
\State Measure the hamming distance of the created code word from the code word of each class.
\State Assign the class to the unknown sample which has the least hamming distance.
\State \textbf{return} Class Label.
\EndProcedure
\end{algorithmic}
\end{algorithm*}

The proposed algorithm including above all methods is described in Algorithm \ref{alecoc}. The EEG dataset contains four different motor task - Thumb movement, Index finger movement, Middle finger movement and Rest. The objective of this study is to classify the recorded data to any one of the desired classes. 80\% of the whole data set were used for training purposes and the remaining 20\%  were used for testing. For this reason, the recorded EEG data after segmented into several sub-bands. First of all, a code matrix was generated using Exhaustive coding method for the desired number of classes. The training process starts with generating two subsets, where the subset $S^{+}$ contains all the positive sample of a particular classifier, and $S^{-}$ contains all the negative samples of that particular classifier according to the code matrix. For more information readers can refer to \cite{dietterich1994solving}. After that, most informative frequency bands were selected from the frequency score. Extra tree binary classifiers were trained to identify the correct class between these two subsets. In the evaluation phase, for an unknown data, first each of the $\lambda_{j}$ were used to predict class label of the sample.Then a code word were generated by joining the class label of each classifier. Hamming distance of the code word were measured by the comparison with the codes of other classes. The data were selected for the class which have lowest hamming distance. 


\section{Result}

\begin{figure*}[htbp]
  \centering
  \includegraphics[width=\textwidth]{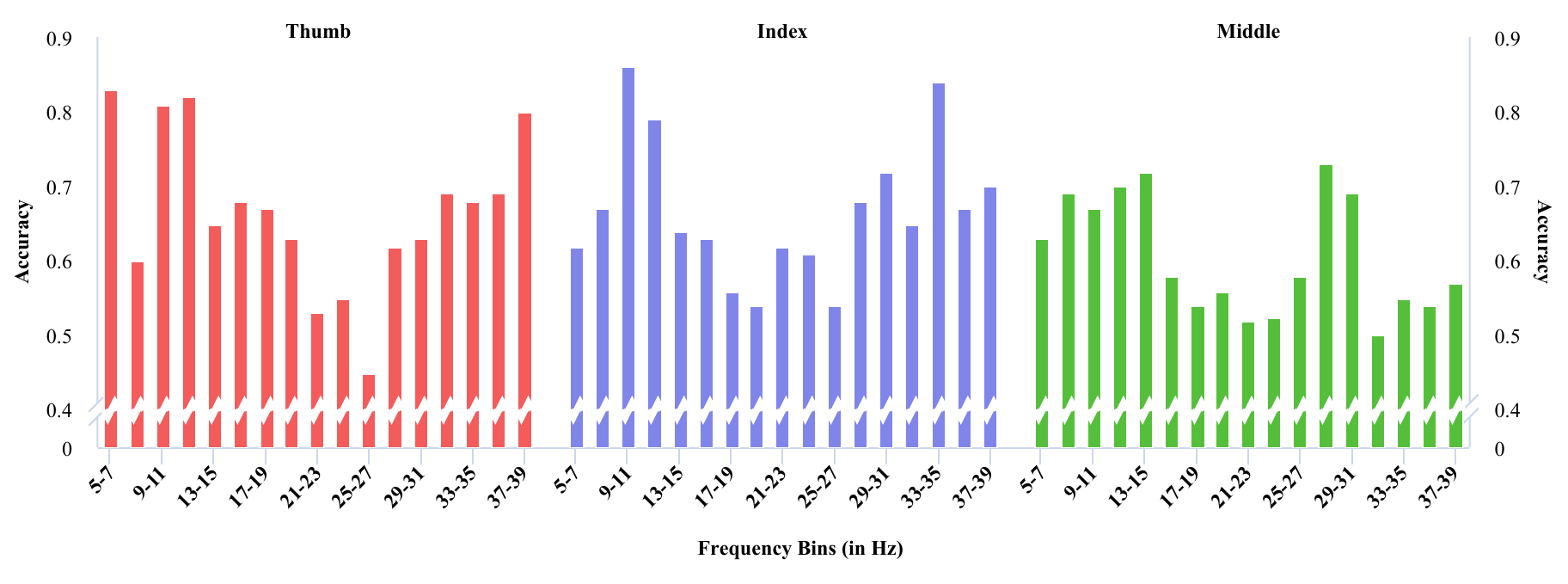}
  \caption{Frequency Scores of Individual Finger}\label{fingerfscore}
\end{figure*}

Figure \ref{fingerfscore} shows the $frequency score$s computed using LDA from the spatially filtered segmented EEG subbands for each finger movements. 
The figure suggests that for the frequency range approximately 9-13 Hz provides the highest $frequency score$ for the thumb as well as index finger movements. 
The frequency bands above a threshold were selected for the classification stage to cope up with inter-subject variations. Table \ref{fbin} tabulates the ranges of selected frequency subbands for each subjects for different finger movement. From that table it can be inferred that the frequency ranges of 8-10 Hz and 10-12 Hz is most common across the subjects. Incidentally, 8-12 Hz frequency range is $\alpha$ band or $\mu$ band. Our results adhered to the fact that it is the most prominent frequency band for finger movement. 


\begin{table*}[htbp]
\caption{Frequency sub band selected by frequency selection algorithm}
\label{fbin}
\centering
\begin{tabular}[scale=0.5]{|c|c|c|c|}
\hline
  & \multicolumn{1}{c|}{\textbf{Thumb}} & \multicolumn{1}{c|}{\textbf{Index}} & \multicolumn{1}{c|}{\textbf{Middle}} \\ \hline
\textbf{S1} & 8-10, 10-12, 12-14 & 8-10, 10-12, 12-14 & 8-10, 10-12, 12-14 \\ \hline
\textbf{S2} & \begin{tabular}[c]{@{}c@{}}4-6, 8-10, 10-12, 12-14\end{tabular} & \begin{tabular}[c]{@{}c@{}}4-6, 6-8, 8-10,10-12, 12-14\end{tabular} & 8-10, 10-12, 12-14 \\ \hline
\textbf{S3} & 4-6 & 4-6, 8-10 & 4-6, 8-10 \\ \hline
\textbf{S4} & 4-6, 8-10, 14-16 & 4-6, 6-8, 8-10, 12-14 & \begin{tabular}[c]{@{}c@{}}4-6, 8-10, 10-12, 20-22\end{tabular} \\ \hline
\textbf{S5} & \begin{tabular}[c]{@{}c@{}}6-8, 8-10, 10-12, 12-14\end{tabular} & \begin{tabular}[c]{@{}c@{}}8-10, 10-12, 12-14, 20-22,\\ 22-24, 24-26
\end{tabular} & \begin{tabular}[c]{@{}c@{}}8-10, 10-12, 12-14, 20-22\end{tabular} \\ \hline
\textbf{S6} & 10-12 & 8-10, 10-12, 12-14 & 8-10, 10-12, 12-14, \\ \hline
\end{tabular}
\end{table*}


The isolated frequency bins were then classified through ET classifier. In this study, the individual finger movements were first classified from the $resting$ state as shown in Table \ref{fingerdecoing}. High classification accuracy for suggests strong contrasts between the spatial patterns of different finger movements and the $resting$ state. For $rest$ vs $Index$ pair, the highest decoding accuracy(DA) was 0.86 surpassing an accuracy with 0.83 for the other pairs.

\begin{table}[htbp]
\caption{Individual finger Decoding Accuracy [Mean$\pm$SD (\textbf{Max})]}
\label{fingerdecoing}
\centering
\begin{tabular}{|c|c|c|c|}
 \hline
\textbf{Subjects} & \textbf{Rest vs Thumb} & \textbf{Rest vs Index} & \textbf{Rest vs Middle} \\ \hline
\textbf{S1} & 0.78$\pm$0.06 (\textbf{0.79}) & 0.72$\pm$0.05 (\textbf{0.79}) & 0.81$\pm$0.03 (\textbf{0.83}) \\
\textbf{S2} & 0.83$\pm$0.02 (\textbf{0.86}) & 0.78$\pm$0.03 (\textbf{0.86}) & 0.74$\pm$0.05 (\textbf{0.86}) \\
\textbf{S3} & 0.59$\pm$0.05 (\textbf{0.62}) & 0.60$\pm$0.03 (\textbf{0.59}) & 0.68$\pm$0.05 (\textbf{0.76}) \\
\textbf{S4} & 0.65$\pm$0.03 (\textbf{0.66}) & 0.81$\pm$0.05 (\textbf{0.86}) & 0.69$\pm$0.05 (\textbf{0.79}) \\
\textbf{S5} & 0.72$\pm$0.03 (\textbf{0.76}) & 0.84$\pm$0.04 (\textbf{0.86}) & 0.83$\pm$0.04 (\textbf{0.83}) \\
\textbf{S6} & 0.76$\pm$0.06 (\textbf{0.83}) & 0.81$\pm$0.03 (\textbf{0.83}) & 0.68$\pm$0.04 (\textbf{0.72}) \\  \hline
\end{tabular}
\end{table}

\begin{table}[]
\caption{Decoding between pair of finger}
\label{fingerpair}
\centering
\begin{tabular}{|c|c|c|c|}
 \hline
\textbf{Subjects} & \textbf{Thumb vs Index} & \textbf{Thumb vs Middle} & \textbf{Index vs Middle} \\ \hline
\textbf{S1} & \begin{tabular}[c]{@{}c@{}}0.59$\pm$0.05 (\textbf{0.76})\end{tabular} & \begin{tabular}[c]{@{}c@{}}0.57$\pm$0.02 (\textbf{0.62})\end{tabular} & \begin{tabular}[c]{@{}c@{}}0.56$\pm$0.04 (\textbf{0.62})\end{tabular} \\ \hline
\textbf{S2} & \begin{tabular}[c]{@{}c@{}}0.67$\pm$0.04 (\textbf{0.68})\end{tabular} & \begin{tabular}[c]{@{}c@{}}0.66$\pm$0.03 (\textbf{0.66})\end{tabular} & \begin{tabular}[c]{@{}c@{}}0.66$\pm$0.04 (\textbf{0.68})\end{tabular} \\ \hline
\textbf{S3} & \begin{tabular}[c]{@{}c@{}}0.56$\pm$0.02 (\textbf{0.59})\end{tabular} & \begin{tabular}[c]{@{}c@{}}0.58$\pm$0.04 (\textbf{0.62})\end{tabular} & \begin{tabular}[c]{@{}c@{}}0.59$\pm$0.02 (\textbf{0.62})\end{tabular} \\ \hline
\textbf{S4} & \begin{tabular}[c]{@{}c@{}}0.57$\pm$0.03 (\textbf{0.59})\end{tabular} & \begin{tabular}[c]{@{}c@{}}0.58$\pm$0.02 (\textbf{0.62})\end{tabular} & \begin{tabular}[c]{@{}c@{}}0.59$\pm$0.02 (\textbf{0.62})\end{tabular} \\ \hline
\textbf{S5} & \begin{tabular}[c]{@{}c@{}}0.58$\pm$0.02 (\textbf{0.62})\end{tabular} & \begin{tabular}[c]{@{}c@{}}0.57$\pm$0.03 (\textbf{0.62})\end{tabular} & \begin{tabular}[c]{@{}c@{}}0.62$\pm$0.02 (\textbf{0.69})\end{tabular} \\ \hline
\textbf{S6} & \begin{tabular}[c]{@{}c@{}}0.67$\pm$0.04 (\textbf{0.72})\end{tabular} & \begin{tabular}[c]{@{}c@{}}0.63$\pm$0.06 (\textbf{0.69})\end{tabular} & \begin{tabular}[c]{@{}c@{}}0.60$\pm$0.05 (\textbf{0.69})\end{tabular} \\  \hline
\end{tabular}
\end{table}

Furthermore, different finger movements were classified in pair at a time to highlight their distinguishability from one another using the same algorithm. Table \ref{fingerpair} shows the pairwise classification accuracy between a pair of fingers. It can be seen that the 0.69 is the highest classification accuracy for the $Thumb$ vs $Index$ pair, whereas it is 0.66 for the other pairs. The binary classification result was used for multi-class classification. The kappa values for classifying between $Thumb$, $Index$, $Middle$ movements, and $resting$ state are shown in Table \ref{fingermulti}. 

\begin{table}[htbp]
\centering
\caption{Multi Class Classification Result for finger movement using REL }
\label{fingermulti}
\begin{tabular}{|l|c|}
\hline
\textbf{Subject} & \textbf{Kappa Value} \\ \hline
S1               & 0.28                 \\ \hline
S2               & 0.29                 \\ \hline
S3               & 0.25                 \\ \hline
S4               & 0.24                 \\ \hline
S5               & 0.36                 \\ \hline
S6               & 0.32                 \\ \hline
\end{tabular}
\end{table}

\section{Discussion}

In this study, an ensemble learning based approach was proposed for individual finger movement identification mutually as well as from $rest$ing position. 
We believe that our study is the first of its kind to identify individual finger movements through a multi-class classifier. In earlier studies, only pairwise identifications i.e.,binary classifications \cite{blankertz2006berlin, liao2014decoding} have been carried out. Most of the literature suggest that the linear classification techniques in EEG are often more suitable than the non-linear one \cite{muller2003linear} due to the problem of over-fitting. The random ensemble learning introduces randomization to avoid this shortcoming. Additionally, it also supersedes necessity of multiplication as in other classification techniques like SVM to considerably reduce computational time. Furthermore, the rule based classification technique makes the ET classifier more faster than others. ECOC technique was summed up here with ET for multi-class decoding ability. The better error correcting ability makes it suitable for BCI use. Due to use of $Exhaustive$ $coding$ technique, maximum hamming distance for each produced code word was 3. So, ECOC can easily correct two errors generated by any two of the binary function.

\begin{figure*}[htbp]
    \centering
    \includegraphics[width=.6\textwidth]{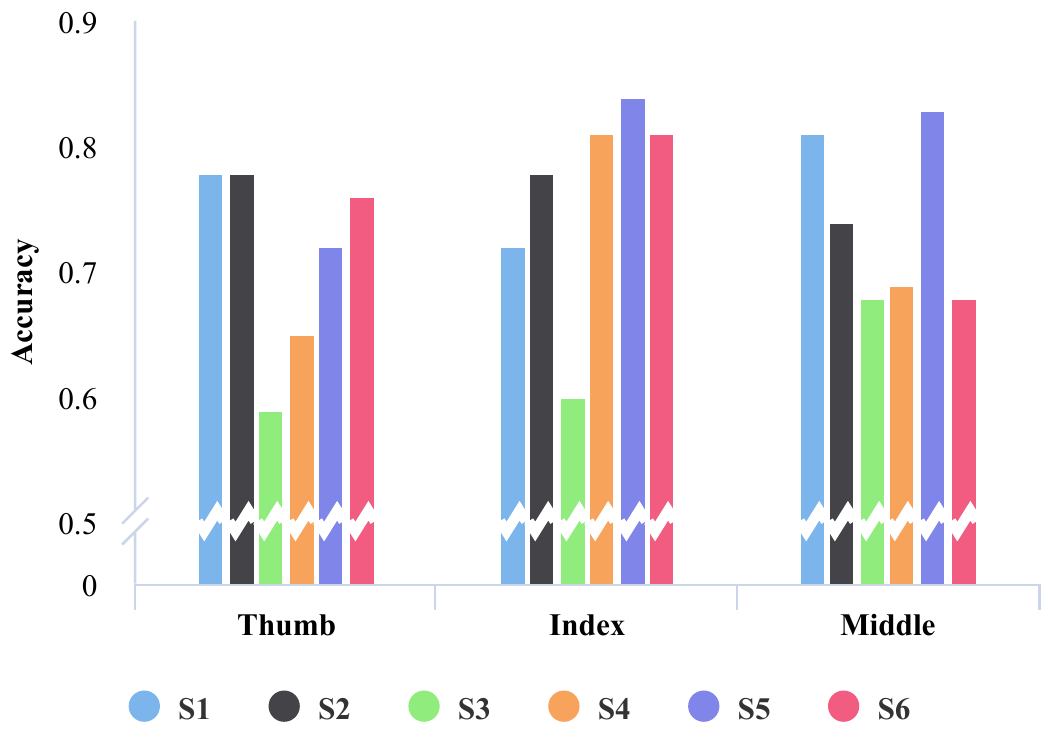}
    \caption{Decoding of Different finger in Different Subjects.}
    \label{finger}
\end{figure*}

\begin{table}[h]
\centering
\caption{Difficulty faced by different subjects while lifting their finger to a certain height.}
\label{fingrd}
\begin{tabular}{|c|c|c|c|}
\hline
\textbf{Subject} & \textbf{Thumb} & \textbf{Index} & \textbf{Middle} \\ \hline
\textbf{S1} & Difficult & Easy & Difficult \\ \hline
\textbf{S2} & Difficult & Difficult & Difficult \\ \hline
\textbf{S3} & Easy & Easy & Easy \\ \hline
\textbf{S4} & Easy & Difficult & Easy \\ \hline
\textbf{S5} & Easy & Difficult & Difficult \\ \hline
\textbf{S6} & Difficult & Difficult & Easy \\ \hline
\end{tabular}
\end{table}

Analyzing the output of the decoding accuracies, it was found that the DA for different finger was different for each subject. There was no such pattern at all. No conclusion can not be made about which finger better DA than other finger. But, at the time of data collection it was observed that few subjects were having difficulties while lifting their some fingers to a certain height. The difficulty faced by different was listed at data collection time for future reference. Table \ref{fingrd} shows the difficulty faced by different subjects. 

If we compare the difficulty faced by different subject with DA of the individual finger from rest condition than we can found that the DA is higher on those finger in which the subject felt problem while lifting his finger. This observation is valid in every subject. In Figure \ref{finger} it can be seen that the decoding of each finger is least in subject 3. Here also we can see that the subject 3 did not felt any problem while lifting his finger. In the case of subject 4, it can be seen that the DA of the index finger is much higher than the other two fingers. Also in subject 6, DA of the middle finger is less than the other two fingers, accordingly subject 6 admitted that he felt difficulty in first two fingers whereas lifting the middle finger is easy for him. With respect to the following observation it can hypothesized that decoding of finger in which the subject feels difficulty at the time of movement has better result. An justification of this hypothesis can be given through the phenomena ERD. According to literature ERD signal represent the mental preparation for any particular motor task \cite{pfurtscheller1999event}. This signal gets reduced in amplitude as the subject masters that specific task. The same is also observed here. A simple intuition can be interpreted, as our brain don't require any attention while going to our own house, wheres going to new place for second or third time requires adequate attention. Additionally, a more widespread ERD occurs in patients and children, as a result the amplitude of ERD increases significantly. \\ 
In Table \ref{fingermulti}, it can be seen that the multi-class DA is low. But it also needs to be noted that, the spatial overlapping of the finger area in motor cortex is very high. So, a 22 channel EEG acquisition is not sufficient to capture the significant difference in individual finger movements. An increase in electrode density may result in higher classification accuracy. Additionally, the concentration of the subjects played an influential role in capturing considerable changes in EEG.
The primary objective of brain computer interface is to enable patient with neuro-muscular disease to communicate with the surrounding. It can also provide smart rehabilitation for post stroke patients. This study confirms that on those cases fine body parts movement can be classified better than of healthy person. The proposed algorithm will be able to assist following application of brain computer interface. 
\section{Conclusion}

This study introduced a new technique for multiclass motor related signal classification. It also investigated competence of the algorithm in discriminating finger movements themselves and also with their $rest$ing state of using noninvasive EEG. The result depicts that algorithm is competent in decoding individuals finger lifting, as well as discriminating between pair of fingers, furthermore results shows that multiclass finger decoding is also possible from noninvasive EEG. The average DA in lifting of individuals finger is 74$\%$ and average DA in discriminating a pair of finger is 60$\%$. 
Furthermore, the endeavor to decode multiclass finger movement from EEG yielded a maximum kappa value of 0.36. As we have used only 22 channels for the recording EEG, with additional channels, DA can be expected to be increased considerably. Our inference suggests that classification accuracy increases in case of difficulty in lifting fingers. As the individuals with neuromuscular disability require more mental preparation to imagine the finger movement, our algorithm will be efficient enough to classify them with high accuracy. This can be utilised in the development of exoskeleton and prostheses with finger control.

\bibliographystyle{IEEEtran}
\bibliography{bibtex/bib/finger}
\end{document}